\documentclass[review]{elsarticle}

\usepackage[T1]{fontenc}
\usepackage{epsfig}
\usepackage{graphicx}
\usepackage{bm}
\usepackage{amsmath}
\usepackage{xcolor}

\usepackage{lineno,hyperref}
\modulolinenumbers[5]

\journal{Spectrochimica Acta A}

\newcommand{\rcm}{\mbox{cm$^{-1}$}}
\newcommand{\Sch}{Schr\"{o}dinger}
\newcommand{\SSst}{$^1\Sigma^{+}_{\mathrm{u}}$}
\newcommand{\SPst}{$^1\Pi_{\mathrm{u}}$}
\newcommand{\TSst}{$^3\Sigma^{+}_{\mathrm{u}}$}
\newcommand{\TPst}{$^3\Pi_{\mathrm{u}}$}

\begin{document}

\begin{frontmatter}

\title{The C(2)$^1\Pi_u$ state in Rb$_2$. Observation and deperturbation.}

\author[Sofia]{Asen Pashov \corref{mycorrespondingauthor}}
\cortext[mycorrespondingauthor]{Corresponding authors}
\ead{pashov@phys.uni-sofia.bg}
\author[FUW]{Pawel Kowalczyk \corref{mycorrespondingauthor}}
\ead{Pawel.Kowalczyk@fuw.edu.pl}
\author[PAN]{Jacek Szczepkowski}
\author[PAN]{Wlodzimierz Jastrzebski \corref{mycorrespondingauthor}} 
\ead{jastr@ifpan.edu.pl}
\address[Sofia]{Faculty of Physics, Sofia University, 5 James Bourchier Boulevard, 1164 Sofia, Bulgaria}
\address[FUW]{Institute of Experimental Physics, Faculty of Physics,
University of Warsaw, ul.~Pasteura~5, 02-093~Warszawa, Poland}
\address[PAN]{Institute of Physics, Polish Academy of Sciences,
al.~Lotnik\'{o}w~32/46, 02-668~Warsaw, Poland}

\date{\today}

\begin{abstract}
We report a systematic study of the C(2)\SPst\ electronic state in rubidium dimer, observed in polarization labelling spectroscopy experiment through the C $\leftarrow$ X$^{1}\Sigma^{+}_{g}$ transitions  recorded under rotational resolution in two isotopologues $^{85}$Rb$_2$ and $^{85}$Rb$^{87}$Rb. Regularity of the vibrational progressions was distorted by numerous interactions with the surrounding 2\SSst, 2\TPst\ and 3\TSst\ states. Deperturbation was performed by coupled-channels analysis taking into account spin-orbit and rotational interactions. Potential energy curves and parameters describing the off-diagonal matrix elements were determined as functions of internuclear distance. About 3000 line frequencies from the present study  are reproduced with a standard deviation of 0.07~\rcm, in agreement with the experimental accuracy. Along with this, the model is consistent with all high resolution experimental data on the involved electronic states, available in the literature. 
\end{abstract}

\begin{keyword}
	laser spectroscopy \sep alkali dimers \sep electronic states \sep
	perturbations \sep deperturbation analysis
	\PACS 31.50.Bc \sep  33.20.Kf \sep 33.20.Vq \sep 33.50.Dq
\end{keyword}

\end{frontmatter}

\date{\today}


\newpage

\section{Introduction}

Diatomic rubidium molecules have attracted much attention in recent years, particularly in view of investigations of ultracold molecules. Consequently, the lowest X$^{1}\Sigma^{+}_{g}$ and a$^{3}\Sigma^{+}_{u}$ states~\cite{Amiot-X,Tsai,Beser,Guan,Seto} and the states correlated with the first excited 5s+5p atomic asymptote, employed in cold physics experiments, have been studied thoroughly (e.g. \cite{Bergeman,Bellos,Takekoshi,Bellos2}). Concerning higher excited electronic states, their spectroscopic investigations by laser techniques started in the last decades of the previous century, in particular from extensive studies by Amiot and co-workers (\cite{Amiot-2Pig,Amiot,Amiot-B}). However, despite of considerable progress achieved in recent years \cite{Drozdova,Arndt,Han,Lee,Lee2,stanD,stany5-1,stany5-2,stany5-3,stany7,stan-dm,stany89,Arndt2,Arndt3}, there are surprising gaps in their experimental knowledge. State-of-the-art theoretical calculations on rubidium dimer \cite{stany5-1,Tomza} give a qualitative picture of its energy structure but precision of the predicted molecular constants is still far below spectroscopic accuracy.

\begin{figure}
	\includegraphics[width=1.0\linewidth]{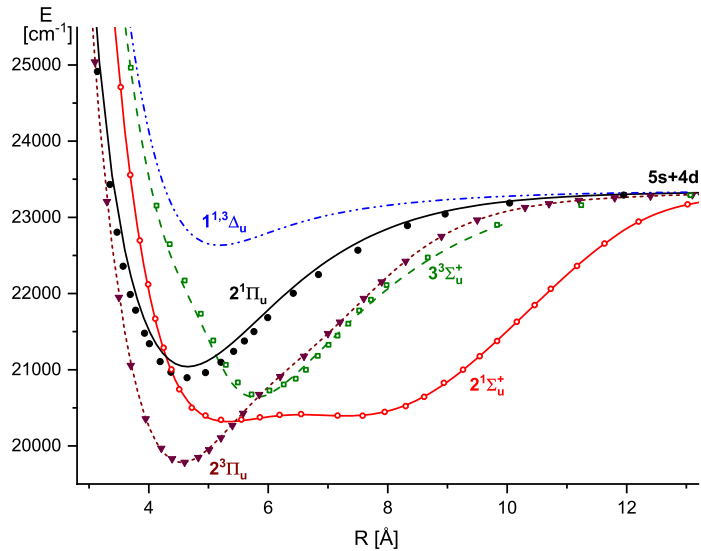}
	\caption{(In colour online) Potential energy curves of electronic states of rubidium dimer relevant for the present analysis. Solid lines indicate singlet states, dashed/dotted lines correspond to states of triplet symmetry as calculated theoretically~\cite{stany5-1}. Circles, squares and triangles represent potentials fitted in the present work. Potential curves of the 1$^1\Delta_{\mathrm{u}}$ and 1$^3\Delta_{\mathrm{u}}$ states (not fitted here) are indistinguishable in this energy scale. Energies are given with respect to the minimum of the ground electronic state. Atomic dissociation products are indicated.}
	\label{potentials}
\end{figure}

In the present paper we are dealing with electronic states correlated with the second excited atomic asymptote in Rb$_2$, 5s+4d, for which the experimental evidence is very scarce and incomplete. The present experiment was originally aimed to study excitation of the 2$^{1}\Sigma^{+}_{u}$ state, according to theoretical calculations located about 20000~cm$^{-1}$ above the bottom of the ground X state and predicted to have a very broad and unusually shaped potential energy curve (see Figure~\ref{potentials}). Up to now only part of this state was observed in a photoassociation experiment \cite{Huang} but not analysed in a spectroscopic sense. Having failed to achieve this goal, possibly because of low transition dipole moment for excitation from low levels of the ground molecular state and/or poor Frank-Condon factors, we focused our attention on the C(2)$^{1}\Pi_{u}$ state, also available in direct transitions from the ground X$^{1}\Sigma^{+}_{g}$ state and positioned in roughly the same energy range. This state was previously investigated in laser-induced fluorescence by Amiot~\cite{Amiot} but the molecular constants determined in that work described only the few lowest vibrational levels and moreover provided low accuracy in reproducing the experimental energies because of evident local perturbations. A limited number of spectral lines observed in the experiment by Amiot did not allow to pursue this problem any further. In later years, the C(2)$^{1}\Pi_{u}$ state was studied again by Lee, Lee and Kim in a cold molecular beam \cite{Lee2}. Their measurements covered broader range of vibrational levels ($v'=0-11$) but were confined to low values of rotational quantum number $J'$. They noticed perturbations of $v'=5-8$ levels and ascribed them to interaction of the C state with the neighbouring 3$^{3}\Sigma^{+}_{u}$ and 2$^{3}\Pi_{u}$ states, whereas the lower $v'=0-4$ and upper $v'=9-11$ levels appeared unperturbed.

In our experiment we recorded systematically excitation spectra of the C(2)$^{1}\Pi_{u}$ $\leftarrow$ X$^{1}\Sigma^{+}_{g}$ band system covering the range $v'=0-25$ and $J'=33-169$ in $^{85}$Rb$_2$ and $^{85}$Rb$^{87}$Rb isotopologues. Perturbations of various strength, resulting from interaction of the C state with 3$^{3}\Sigma^{+}_{u}$ and 2$^{3}\Pi_{u}$ but also with the 2$^{1}\Sigma^{+}_{u}$ state (the last perturbation previously undetected) have been found for all the observed levels, even those claimed to be free of perturbation in previous reports. The three perturbing states turned out to be directly unobservable (`dark' states) except for close vicinity to regions where perturbations culminate. After careful examination of the spectra finally we have assigned nearly 3000 spectral lines of the C(2)$^{1}\Pi_{u}$ $\leftarrow$ X$^{1}\Sigma^{+}_{g}$ system, including both main lines and extra lines. 

All the existing experimental data, the vast majority of them collected in the present experiment and supplemented by results of previous reports \cite{Lee,Lee2}, were subjected to coupled-channels deperturbation analysis outlined below, in which potential energy curves of all four involved electronic states and parameters describing couplings between them were determined as functions of internuclear distance. The model allows to reproduce the experimental positions of spectral lines with an overall accuracy of 0.07~cm$^{-1}$.

\section{Experimental procedure}
\label{Exp_proc}

The experimental principle and setup are very similar to those described in the previous papers \cite{stanD,stany5-1,stany5-2,stany5-3,stany7,stan-dm,stany89}. Laser polarization labelling spectroscopy (PLS) was employed to study excitation spectra of Rb$_2$ from the electronic ground state. For this purpose the molecular sample was irradiated by co-propagating beams of two independent lasers, the strong, tuneable pump laser and weaker, narrowband probe laser. The frequency of the probe laser was set on known molecular transitions in the B(1)$^{1}\Pi_{u}$ $\leftarrow$ X$^{1}\Sigma^{+}_{g}$ band system of rubidium dimer \cite{Amiot-B}. Light of the pump laser was tuned across the investigated spectral region. Whenever its frequency coincided with some molecular transition ($v'$,$J'$) $\leftarrow$ ($v''$,$J''$), an optical anisotropy was induced in the rovibrational level of the ground state ($v''$,$J''$) by populating or depleting the degenerate $M_J''$ levels to various degrees. As the linearly polarized probe beam interacted with this anisotropic sample, its polarization was changed, provided that the lower level ($v''$,$J''$) was shared by the pump and probe transitions. This was monitored by two crossed polarizers placed on opposite sides of the molecular sample. Thus fixing the probe laser on a chosen, well known B($v$,$J$) $\leftarrow$ X($v''$,$J''$) transition, tuning the pump laser and observing changes in polarization of the probe beam we could record transitions from a single (`labelled') ground state level to many levels of the excited state(s), forming progressions of P, R doublets or P, Q, R triplets depending on polarization of the pump light (see Ref.~\cite{PLS} for detailed selection rules applying to the PLS method). 

The rubidium dimers were produced in a linear heat-pipe oven by heating metallic rubidium (ca.~5~g, natural isotopic abundance) to about 570~K in presence of 4~Torr of argon buffer gas protecting the quartz windows. As a probe laser we used a cw single-mode ring dye laser (Coherent 899-21 operated on DCM dye) set at fixed frequencies of the B $\leftarrow$ X system measured and actively stabilized using a High-Finesse WS-7 wavemeter. The pump laser (pulsed Lumonics HD 500 dye laser) with Coumarin 480 as a laser dye was scanned between 20400 and 21350~cm$^{-1}$, allowing molecular lines originating from the labelled ground state levels to be recorded in this range. With the laser bandwidth slightly below 0.1~cm$^{-1}$, the spectra were calibrated to a similar absolute accuracy, accomplished by simultaneous recording of argon and neon optogalvanic lines as well as transmission fringes of a Fabry-P\'{e}rot etalon 0.5~cm long. 

\section{Experimental observations}
\label{Exp_obs}

\begin{figure}
	\includegraphics[width=1.1\linewidth]{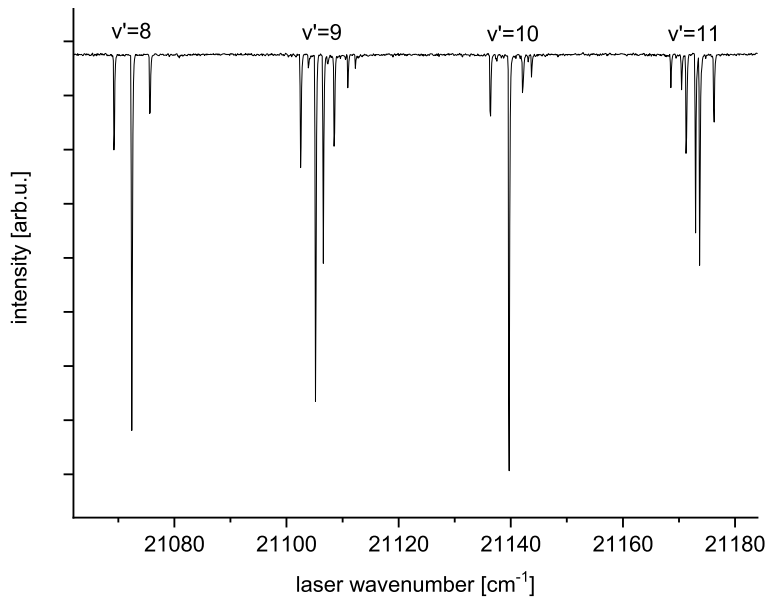}
	\caption{A portion of the experimental spectrum corresponding to the C(2)$^{1}\Pi_{u}$($v',J'=J'',J'' \pm 1$) $\leftarrow$ X$^{1}\Sigma^{+}_{g}$($v''=1,J''=91$) transition originating from the ground state level of $^{85}$Rb$_2$ labelled by the probe laser set at the known B(1)$^{1}\Pi_{u}$($v=6,J=91$) $\leftarrow$ X$^{1}\Sigma^{+}_{g}$($v''=1,J''=91$) transition at the wavenumber 14854.515~cm$^{-1}$. Only the $v'=8$ level of the C state appears unperturbed as at excitation  of the other levels some extra lines can be seen.}
	\label{spectrum}
\end{figure}

The investigated spectral region corresponds to the energy range 19800--22100~cm$^{-1}$ above the minimum of the ground state potential well. We recorded there transitions from the  X$^{1}\Sigma^{+}_{g}$ state apparently to one excited electronic state. Symmetry of the upper state has been established as $^{1}\Pi_{u}$ because of presence of Q lines in the spectra and it has been identified as C(2)$^{1}\Pi_{u}$ basing on molecular constants given by Amiot~\cite{Amiot}. However, the C state turned out to be strongly perturbed, with corresponding spectral lines frequently shifted even by a few cm$^{-1}$ from the positions predicted by the existing constants. In addition, numerous extra lines accompanied the main lines (Figure~\ref{spectrum}). 

In Figure~\ref{1267}a we display differences between the observed energy levels and the levels calculated for $v'=1,2,6$ and 7 in a simple single channel model of the C(2)\SPst\ state. These vibrational levels are selected to show two different types of perturbations. In $v'=1,2$ only the $e$ symmetry levels are affected and frequent culminations of perturbations tell us that the vibrational and rotational constants of the perturber should be very different from that of the C state. Comparing Fig.~\ref{potentials} one can guess that the perturber is in this case the 2\SSst\ state. In $v'=6,7$ levels of both symmetries are affected and fewer culminations of perturbations are observed, so the perturbers can be 2\TPst\ and 3\TSst\ states. 

Altogether 2978 main and extra lines were identified in the spectra of both isotopologues $^{85}$Rb$_2$ and $^{85}$Rb$^{87}$Rb, and assigned $v'$ and $J'$ quantum numbers of the upper state when it was dominantly the C state, and only $J'$ in case of perturbing states. The calibration of the PLS spectra assures uncertainty better than 0.1~\rcm. In the course of further analysis only 28 lines (i.e. less than 1\% of all observed) have been rejected as misassigned or impossible to be included in the model developed below since they cannot be reproduced within better than $\pm 0.5$ \rcm. The measured line positions are differences between the energies of the upper and lower levels. Since the energies of rovibrational levels in the ground X$^{1}\Sigma^{+}_{g}$ state are known with very high precision \cite{Seto}, the upper state energies were determined with accuracy of the spectral line measurements. This resulted in 2035 energies of the upper state(s) levels. As our observations span the range $v'=0-25$, $J'=33-169$ in the C(2)$^{1}\Pi_{u}$ state, we supplemented the database by adding energies of levels generated from band parameters given by Lee \textit{et al.} in Table 1 of Ref.~\cite{Lee2} and related to the lowest rotational levels $J' \textless 32$ for $v'=0-11$. We added also some energy levels of the 2\TSst\ state ($J'=1-10,  v'=10-13$, constants from Table 2 in Ref.~\cite{Lee2}) and of two $\Omega$ components of the 2\TPst\ state, taken from Table I ($J'=1-9, v'=0-12$) and Table II ($J'=1, v'=0-12$) of Ref.~\cite{Lee}, generated in the same way. Consequently, a total of 542 calculated energy levels from \cite{Lee,Lee2} were added. The range of rotational quantum numbers $J$ of the added levels is somewhat arbitrary, because in \cite{Lee,Lee2} there is no information about the validity of the derived band constants.

\section{The coupled-channels (CC) model}
\label{CC_mod}

For reporting the experimental observations we attempted to construct as simple and compact model as possible. The analysis started with the $f$ symmetry levels, observed through the Q transitions. Such transitions initially dominated the experimental data because up to $v'=4$ level they seemed to be free from local perturbations. Based on these data it was found that the main perturber of the C(2)\SPst\ state is not the 2\TPst\ state (as stated by Lee \textit{et al.}~\cite{Lee2}), but rather the 3\TSst. In order to model interactions between the C(2)\SPst\ and the 3\TSst\ states we need already a three channels model including the $\Omega=0^-$ and $\Omega=1$ components of the 3\TSst\ state. The $\Omega=1$ component of the 2\TPst\ state also perturbs the C(2)\SPst\, but mostly indirectly, through its interaction with the 3\TSst. This four channels model was quite adequate for the $f$ symmetry levels. To extend it to the $e$ symmetry manifold we removed the $3^3\Sigma^+_{0^-}$ channel ($f$ symmetry only), but it turned out necessary to add the 2\SSst\ state, because the $e$ symmetry levels were found to be perturbed already starting from $v'=0$ and interaction with the 2\SSst\ state ($e$ symmetry only) is the only plausible reason for this. Such an interaction was not reported in the previous study by Lee \textit{et al.}~\cite{Lee2}, probably because of low $J'$ levels observed in their beam experiment. Finally, the four channels model for both symmetries was again extended by including the $\Omega=0^{\pm}$ components of the 2\TPst\ state, because it turned out to be important to reduce the excessively large residuals for the heavily mixed \SPst $\sim$ \TSst\ levels, especially of $f$ symmetry. With the present set of experimental data we see no necessity to include the $\Omega=2$ component of the 2\TPst\ state or the 1$^{1,3}\Delta_{\mathrm{u}}$ states which correlate to the same 5s+5d asymptote, so the final model for both symmetries contains five channels. 

In the formulation outlined above, the model involves four electronic states, which are coupled by the spin-orbit and various rotational perturbations. We aimed to include only the most significant channels, sufficient to reproduce the experimental observations. In other words we took into account only these channels, removing of which degraded quality of the fit significantly. However, the complexity of the model and structure of the experimental data (containing mainly data on the C state) caused that some components of the model could not be determined uniquely. Fortunately our task was facilitated by high quality {\it ab initio} calculations providing theoretical potential energy curves \cite{stany5-1} and various elements of the Hamiltonian matrix linking the involved states \cite{Pazyuk:2015}.

In the present section we summarize the components of the CC model. Since we use symmetrized electronic wave functions while in \cite{Pazyuk:2015} the matrix elements are calculated with nonsymmetrized functions, it is important to show explicitly the expressions for matrix elements used by us in order to avoid possible ambiguities.

In the model we have six spin-orbit interactions (for sake of simplicity hereafter we omit the ungerade index and plus/minus indices wherever no ambiguity is introduced)

\begin{equation} 
\begin{split}
\langle^1\Pi_1|\hat{H}^{\mathrm{SO}}|^3\Sigma^+_1\rangle_{ef} &= \xi_{^1\Pi^3\Sigma}(R)\\
\langle^1\Pi_1|\hat{H}^{\mathrm{SO}}|^3\Pi_1\rangle_{ef} &= -\xi_{^1\Pi^3\Pi}(R)\\
\langle^3\Sigma^+_1|\hat{H}^{\mathrm{SO}}|^3\Pi_1\rangle_{ef} &= \xi_{^3\Sigma^3\Pi}(R)\\
\langle^3\Sigma^+_0|\hat{H}^{\mathrm{SO}}|^3\Pi_0\rangle_{f} &= \sqrt{2}\xi_{^3\Sigma^3\Pi}(R)  \\
\langle^1\Sigma^+_0|\hat{H}^{\mathrm{SO}}|^3\Pi_0\rangle_{e} &= -\sqrt{2}\xi_{^1\Sigma^3\Pi}(R) \\
\langle^3\Pi_0|\hat{H}^{\mathrm{SO}}|^3\Pi_0\rangle_{ef} &= -A_{^3\Pi^3\Pi}(R)\mbox{ ,}\\
\end{split}
\label{SOs}
\end{equation}  

\noindent and seven rotational interactions. The latter are caused by the neglected part of the rotational Hamiltonian $\hat{H}^{\mathrm{rot}}=B\big(L^\pm S^\mp-J^\pm S^\mp - J^\pm L^\mp\big)$. Here $J^\pm L^\mp$ stands for $J^+L^-+J^-L^+$ and $B=\hbar^2/(2mR^2)$.

\begin{equation} 
\begin{split}
J^\pm S^\mp\mathrm{coupling:}&\\
\langle^3\Sigma^+_1|\hat{H}^{\mathrm{rot}}|^3\Sigma^+_0\rangle_{f} &= -B\sqrt{2}\sqrt{S(S+1)}\sqrt{J(J+1)}\\
\langle^3\Pi_1|\hat{H}^{\mathrm{rot}}|^3\Pi_0\rangle_{ef} &= -B\sqrt{S(S+1)}\sqrt{J(J+1)}\\
L^\pm S^\mp\mathrm{coupling:}&\\
\langle^3\Sigma^+_1|\hat{H}^{\mathrm{rot}}|^3\Pi_1\rangle_{ef} &= B\sqrt{S(S+1)}L_{^3\Pi^3\Sigma}(R)\\
\langle^3\Sigma^+_0|\hat{H}^{\mathrm{rot}}|^3\Pi_0\rangle_{f} &= B\sqrt{2}\sqrt{S(S+1)}L_{^3\Pi^3\Sigma}(R) \\
J^\pm L^\mp\mathrm{coupling:}&\\
\langle^1\Pi_1|\hat{H}^{\mathrm{rot}}|^1\Sigma^+_0\rangle_{e} &= -B\sqrt{2}\sqrt{J(J+1)}L_{^1\Pi^1\Sigma}(R) \\
\langle^3\Sigma^+_1|\hat{H}^{\mathrm{rot}}|^3\Pi_0\rangle_{ef} &= -B\sqrt{J(J+1)}L_{^3\Pi^3\Sigma}(R)\\ 	
\langle^3\Sigma^+_0|\hat{H}^{\mathrm{rot}}|^3\Pi_1\rangle_{f} &= -B\sqrt{2}\sqrt{J(J+1)}L_{^3\Pi^3\Sigma}(R)\mbox{ .}\\
\end{split}
\label{rots}
\end{equation} 

\noindent The lower indices $e$ and $f$ of the matrix elements denote for which symmetry manifold they are calculated. Generally the calculation of matrix elements follows Ref.~\cite{Kato:1993}, the only differences are in the matrix elements involving the 3\TSst\ state. In \cite{Kato:1993} the $S$-uncoupling operator ($-BJ^\pm S^\mp$) is taken explicitly into account and the wave functions are actually the Hund's case (b) functions. Here the 3\TSst\ state is treated within the Hund's case (a).

Details on the realization of the CC model can be found in Refs.~\cite{stany5-3,Szczepkowski:2019}. The total wave function is represented in the basis of the selected Hund's case (a) states $\Phi_i(R,r)$ as:

\begin{equation}
\Psi_\alpha(R,r)=\sum^N_{i=1}\Phi_{i\alpha}(R,r)\chi_{i\alpha}(R)\mbox{ ,}
\label{wavefun}
\end{equation}

\noindent where $\chi_{i\alpha}(R)$ are $R$-dependent mixing coefficients. Index $\alpha$ denotes all good quantum numbers, which characterize the mixed state ($J$, $e/f$ symmetry). For the $e$ symmetry levels the basis functions $\Phi_{i\alpha}(R,r)$ are $2^1\Pi_1$,  $2^1\Sigma^+_0$, $3^3\Sigma^+_1$, $2^3\Pi_1$ and $2^3\Pi_0$. For the $f$ symmetry levels the basis functions are  $2^1\Pi_1$,  $3^3\Sigma^+_0$, $3^3\Sigma^+_1$, $2^3\Pi_1$ and $2^3\Pi_0$. Coefficients $\chi_{i\alpha}(R)$ are searched as solutions of the system of coupled \Sch\ equations

\begin{equation}
\Big(-\frac{\hbar^2}{2mR^2}\frac{d^2}{dR^2}+U_{i}(R)+\hat{H}_{i\alpha}^{\mathrm{ROT}}(R)\Big)\chi_{i\alpha}(R)+\sum_{j \neq i}\hat{H}_{ij}(R)\chi_{j\alpha}=E_{\alpha}\chi_{i\alpha}(R) \mbox{ ,} 
\label{1}
\end{equation}

\noindent where $U_{i}(R)$ is the diagonal $R$-dependent matrix element of the electronic Hamiltonian. It contains the rotationless potential energy curve for each of the basis electronic states and also diagonal matrix elements of the spin-orbit operator. $\hat{H}_{i\alpha}^{\mathrm{ROT}}(R)$ denotes the diagonal angular part of the nuclear kinetic energy operator

\begin{equation}
\hat{H}_{i\alpha}^{\mathrm{ROT}}=\frac{\hbar^2}{2mR^2}\big(J(J+1)-\Omega^2-\Lambda^2+S(S+1)-\Sigma^2\big)\mbox{ ,}
\end{equation}

\noindent and $\hat{H}_{ij}(R)$ represent the off-diagonal matrix elements of the spin-orbit operators (\ref{SOs}) and of the rotational Hamiltonian~(\ref{rots}).

The system of CC equations is solved by the Fourier-Grid-Hamiltonian method \cite{FGH} implemented as described earlier \cite{stany5-3}. All radial functions are calculated in a mesh of 250 points for $2.5\leq R\leq 14 $ \AA. Such a relatively large interval is necessary due to the double minimum 2\SSst\ state, the classical outer turning points of which are at about $10-11$ \AA\ for the studied range of energies. The matrix of the model Hamiltonian is then diagonalized and for each experimental level $E^{\mathrm{exp}}$ the closest match among the calculated energies $E^{\mathrm{calc}}$ is searched. The model parameters are then fitted to minimize value of the expression

\[\bar{\sigma}=\sqrt{\frac{1}{n}\sum_i\frac{(E^{\mathrm{exp}}_i-E^{\mathrm{calc}}_i)^2}{\sigma_i^2}} \mbox{ .}\]

All model functions are defined as sets of points connected with cubic splines. The ordinates of these points are adjusted during the fit. The model consists of four PECs, two $R$-dependent matrix elements of the $L^+$ operator ($L_{^1\Pi^1\Sigma}(R)$ and $L_{^3\Pi^3\Sigma}(R)$) and five SO matrix elements: the off-diagonal $\xi_{^1\Pi^3\Sigma}(R)$, $\xi_{^1\Pi^3\Pi}(R)$, $\xi_{^3\Sigma^3\Pi}(R)$ and $\xi_{^1\Sigma^3\Pi}(R)$ and the diagonal $A_{^3\Pi^3\Pi}(R)$. Initially the matrix elements of the coupling operators were taken from Ref.~\cite{Pazyuk:2015}, but the number of grid points was reduced to about $10$--$15$. At later stages of the analysis some of these functions were fitted.

The most sensitive and ambiguous part of the model are the PECs. Before the current study systematic experimental observations existed only for the C(2)\SPst\ ~\cite{Lee2} and 3\TPst\ states~\cite{Lee}. They allow to fix the $T_e$ values of these electronic states and also to some extent the shape of their PECs (by the RKR potentials derived in \cite{Lee} and \cite{Lee2}), up to $v'=11$ in the C state and $v'=12$ in the 2\TPst\ state. For longer and shorter internuclear distances we extended the RKR potentials by theoretical curves from Ref.~\cite{stany5-1}.

Four bands of the 3\TSst--X$^1\Sigma^+_{\mathrm{g}}$ system were also observed in \cite{Lee2} and characterized by their band origins $\nu_0$, rotational constants $B_{v'}$ and the $\Lambda$-doubling constants. Isotope analysis provided the most likely vibrational quantum numbers of these bands as $v'=10-13$. This information was used by us to adjust appropriately the theoretical potential curve of the 3\TSst\ from Ref.~\cite{stany5-1}. As already mentioned, this state turned out to be the most strong and direct perturber of the C state, however at the early stages of the analysis it was found that similar quality of the fit may be achieved using different values of $T_e$ for 3\TSst\ ($\approx \pm 100$ \rcm, while $\omega_e$ is about 40 \rcm). The data from Ref.~\cite{Lee2} gave ultimate preference to one of these values. Eventually, the band constants for the C(2)\SPst, 2\TPst\ and 3\TSst\ states from \cite{Lee, Lee2} were used to calculate `experimental' energies of the corresponding levels and these energies were added to the present experimental set of data. 

\begin{figure}
	\includegraphics[width=1.0\linewidth]{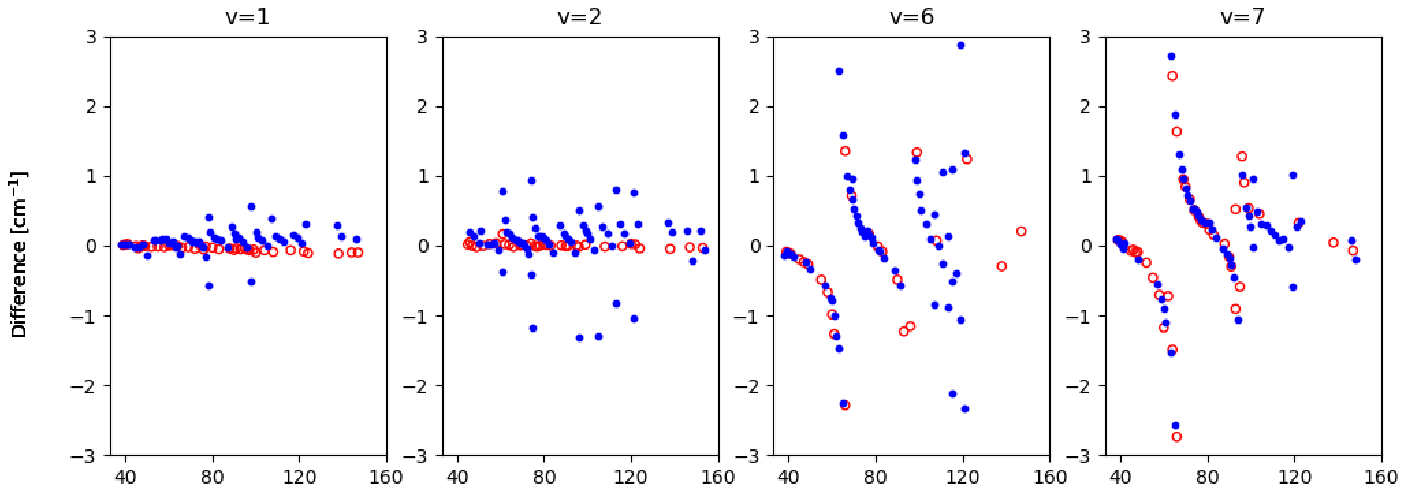}a
	\includegraphics[width=1.0\linewidth]{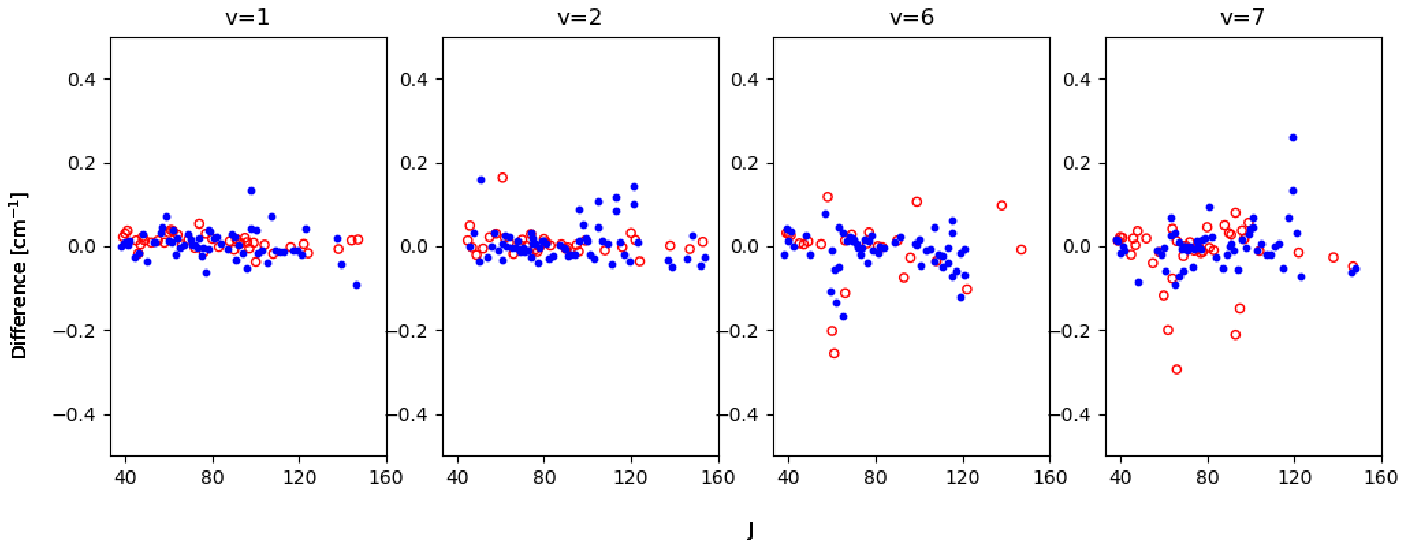}b
	\caption{Differences between the observed and calculated energy levels for $v'=1,2,6$ and 7. In panel (a) the levels are calculated from a single channel model, in panel (b) from the present CC model. Full circles (blue online) correspond to $e$ parity levels, open circles (red online) to $f$ levels. The vibrational levels $v'$ are selected to show various types of perturbations in the 2\SPst\ state. Note different vertical scales in both panels.}
	\label{1267}
\end{figure}

The only state which so far avoided experimental characterization at high resolution is the 2\SSst\ state. Several attempts were undertaken within the present study but no PLS spectra related to this state were registered for transitions from the ground state. At present the reason remains unclear, since an estimation of relative line intensities based on the transition dipole moment from Ref.~\cite{Tomza} and the corresponding overlap integrals show that transitions in the 2\SSst\ $\leftarrow$ X$^1\Sigma^+_{\mathrm{g}}$ band system should not be very much weaker than in the C(2)\SPst\ $\leftarrow$ X$^1\Sigma^+_{ \mathrm{g}}$ system observed by us. The 2\SSst\ state is responsible mainly for the $J$-dependent, heterogenous perturbations with $e$ levels of the C state near its minimum. These perturbations become clearly visible for $J'>30$ and therefore most of them were not noticed in the earlier study by Lee \textit{et al.}~\cite{Lee2}. Due to a large difference in the rotational constants of the interacting states, for each vibrational quantum number (up to $v'=6-7$) level crossings appear approximately for every tenth rotational quantum number $J$ (see Fig.~\ref{1267}a). Again, these perturbations could be reasonably reproduced with various positions of the 2\SSst\ state potential minimum. To reduce the ambiguity, we carried out systematic measurements covering a broad range of rotational quantum numbers $61 \leq J \leq 121$ in the main isotopologue $^{85}$Rb$_2$ and the second abundant isotopologue $^{85}$Rb$^{87}$Rb. The isotope shift depends on the depth of the potential well and by having significant amount of data on both isotopologues we found the unique value of $T_e$ of the 2\SSst\ state which is able to reproduce correctly perturbations in both isotopologues.

\section{Results}
\label{Res}

As already mentioned in Section~\ref{CC_mod}, the complexity of the model grew as more and more experimental data were involved. The ambiguity in the necessary number of channels originates mainly from the structure of our data set, containing predominantly C(2)\SPst\ levels. Although including many perturbed energy levels, these data did not allow to construct the potential curves and the coupling matrix elements in a unique way. During the fit, we often observed that very different potential curves for the 3\TSst, 2\TPst\ and 2\SSst\ states led to the same results. Similar observation concerns the shape of the radial functions $\xi(R), A(R), L(R)$, included in the coupling matrix elements. Therefore we cannot claim that the final model functions, resulting from numerous trials and errors, are unique. Very helpful in removing ambiguities were the existing data on the 3\TSst\ and 2\TPst\ states from \cite{Lee,Lee2} as well as the theoretical curves from \cite{stany5-1} and \cite{Pazyuk:2015}. 

The final parameters for the best model functions are listed in the supplementary materials \cite{EPAPS}. In Figure~\ref{potentials} we compare the fitted potentials with the theoretical ones from Ref.~\cite{stany5-1}. Significant difference can be observed only for the C(2)\SPst\ state, but this disagreement was noticed already in \cite{stany5-1} by comparison with the experimental observations by Amiot \textit{et al.}~\cite{Amiot}. The present study estimates the minimum of the C state potential curve to be at $R_e=4.645$~\AA\ with $T_e=20895.16(2)$ \rcm, in agreement with Refs. \cite{Amiot,Lee2}. In the scale of the Figure~\ref{potentials} one can hardly see differences between the other fitted and theoretical potential curves and indeed they agree to within approximately $\pm 100$ \rcm. For the 2\TPst\ state, the minimum of the PEC is fixed by the experimental data from Ref.~\cite{Lee}. However for the other two potentials, that is of the 3\TSst\ and 2\SSst\ states, the close agreement results rather from our intention to keep the fitted curves as close to the theoretical ones as possible, than from fitting of the experimental data. The ambiguity in this case may be removed only by new experimental observations.

Apart from the potential curves, the model contains altogether seven radial functions (see eqs.~(\ref{SOs}) and (\ref{rots})). Initially they were all fixed to the theoretical values from Ref.~\cite{Pazyuk:2015}. Already the first iterations showed that the most important functions in the fit are the PECs of the C and the 3\TSst\ states and the $\xi_{^1\Pi^3\Sigma}(R)$ radial function. They had to be adjusted virtually always when new experimental data were added or changes in the model introduced. The potential curve of the 2\SSst\ state was adjusted at the beginning, when solely the $e$ symmetry levels were included in the fit and later was readjusted only at the final stages of the fit. The apparent reason is that the C(2)\SPst\ $\sim$ 2\SSst\ interaction can be to a great extent separated from the rest of the model. To describe this interaction within the experimental uncertainty, the radial function $L_{^1\Pi^1\Sigma}$ was also adjusted together with the 2\SSst\ PEC. Surprisingly, the PEC of the 2\TPst\ needed much less adjustment, although significant number of the observed levels contains admixture of this state. 

\begin{figure}
\centering
	\includegraphics[width=0.7\linewidth]{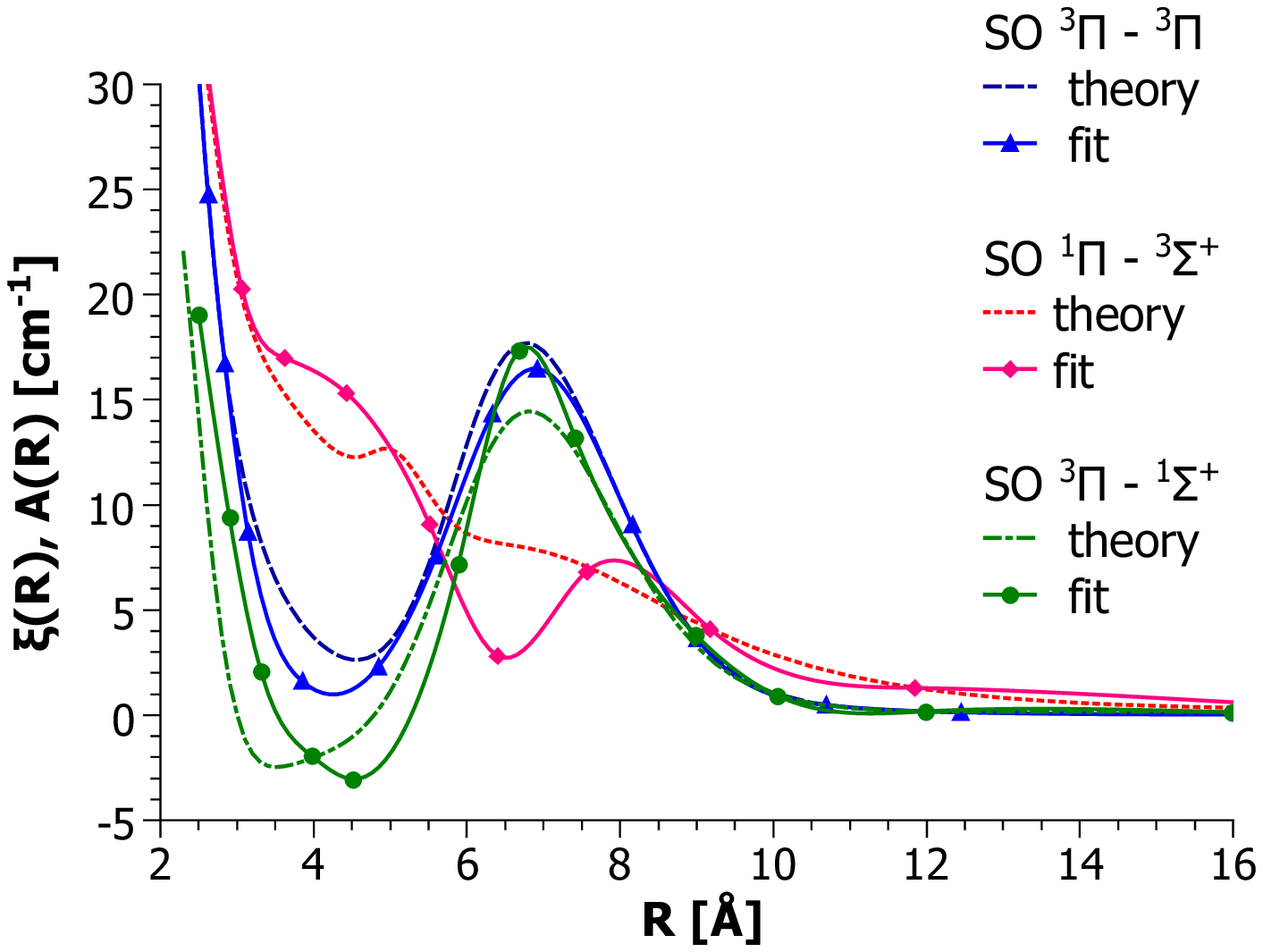}a
	\includegraphics[width=0.7\linewidth]{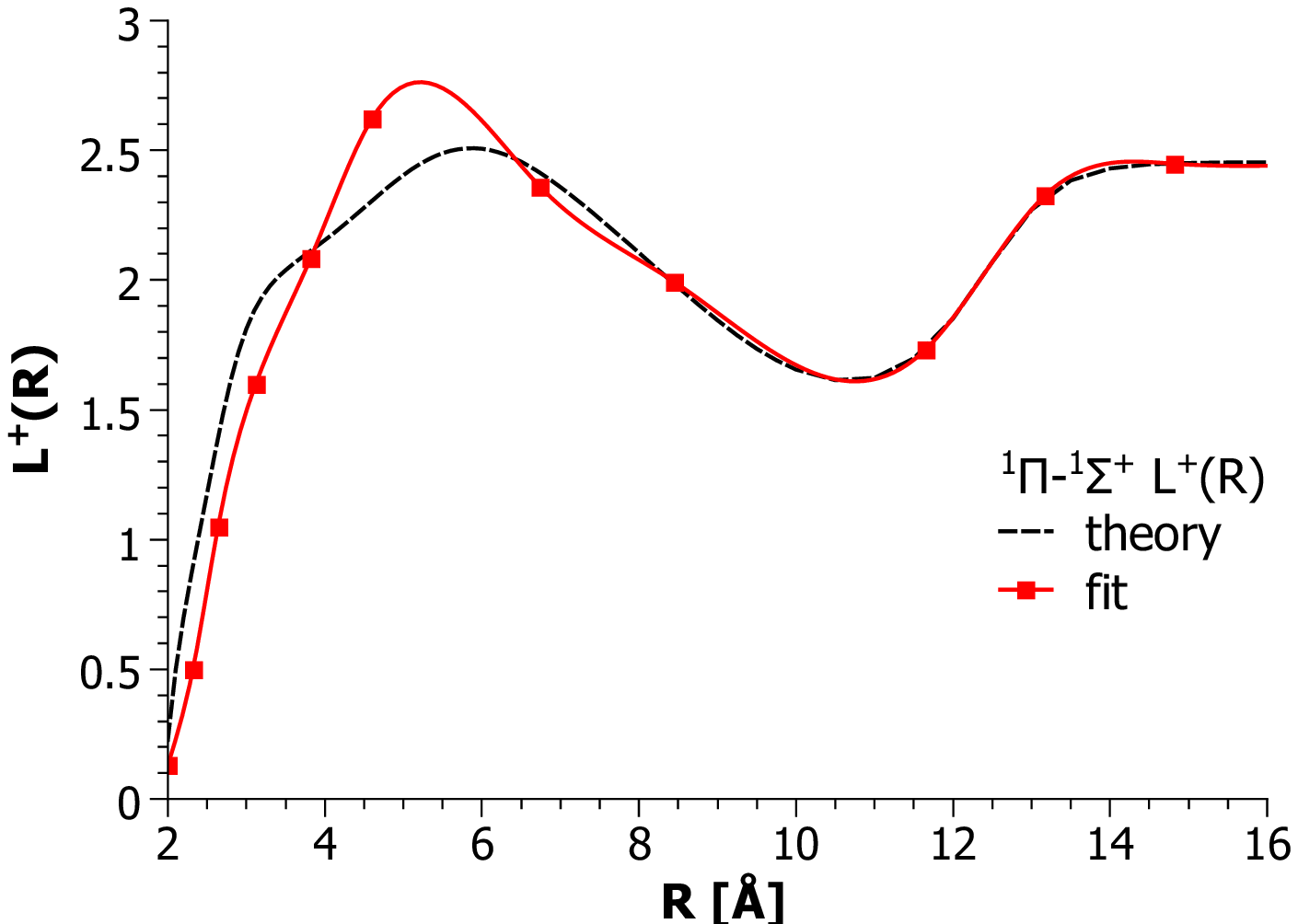}b
	\caption{Radial functions of selected matrix elements, adjusted in the fit, compared with the theoretical values~\cite{Pazyuk:2015}. The other radial functions (see eq.~\ref{SOs} and \ref{rots}) are the same as presented in Ref.~\cite{Pazyuk:2015}.}
	\label{functions}
\end{figure}

At the final stages of the analysis, a specific feature of the problem was that most of the experimental data could be reasonably reproduced by a four channels model and only levels in close vicinity of the avoided crossings between the rovibrational ladders of the interacting states showed significant disagreement. While the residuals for the $e$ symmetry levels were distributed convincingly around zero within the experimental uncertainty, some of the $f$ symmetry levels deviated from the predictions of the model around culminations of the C(2)\SPst\ $\sim$3\TSst\ perturbation. Therefore the $\Omega=0^{\pm}$ components of the 2\TPst\ were subsequently added to the model. The 2$^3\Pi_{0^-}$ and 2$^3\Pi_{0^+}$ components have the same PECs as the 2$^3\Pi_1$ one but they differ by the diagonal spin-orbit matrix element $A_{^3\Pi^3\Pi}(R)$ (see eq.~\ref{SOs}), which was adjusted to achieve the desired agreement in the $f$ symmetry manifold. However, the corresponding new channel (2$^3\Pi_{0^+}$) in the $e$ manifold led to slight degradation of the fit quality in the region of strong C(2)\SPst\ $\sim$ 3\TSst\ mixing for $v'=4-6$. The model for the $e$ and the $f$ symmetry levels looks very similar except for the 2\SSst\ state, present only in the Hamiltonian matrix for $e$ parity levels and for the 3$^3\Sigma^+_{0^-}$ state, present only in the matrix for $f$ levels. Therefore it was suspected that interaction between the 2$^3\Pi_{0^+}$ and 2$^1\Sigma^+_{0}$ states caused the problems. Indeed, a slight adjustment of the $\xi_{^1\Sigma^3\Pi}(R)$ functions restored the fit quality. The finally fitted coupling functions are shown in Figure~\ref{functions} and the rest of them are taken without changes from Ref.~\cite{Pazyuk:2015}. All radial functions and PECs of the model are listed in the Supplementary materials \cite{EPAPS}.

\begin{figure}
	\includegraphics[width=1.0\linewidth]{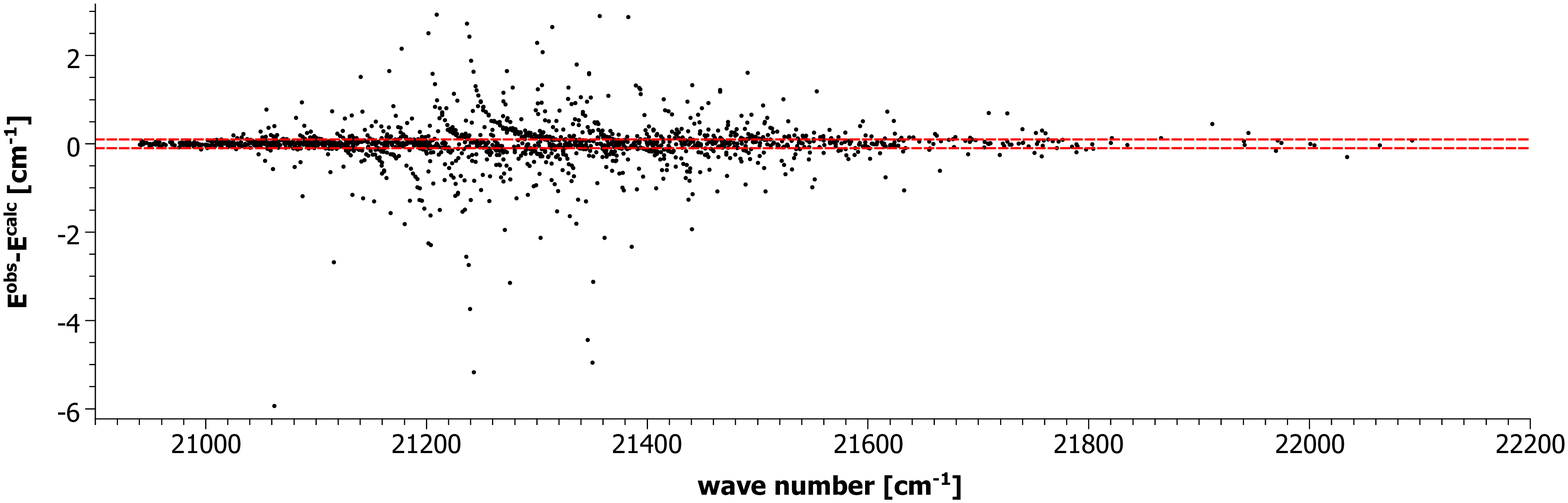}a
	\includegraphics[width=1.0\linewidth]{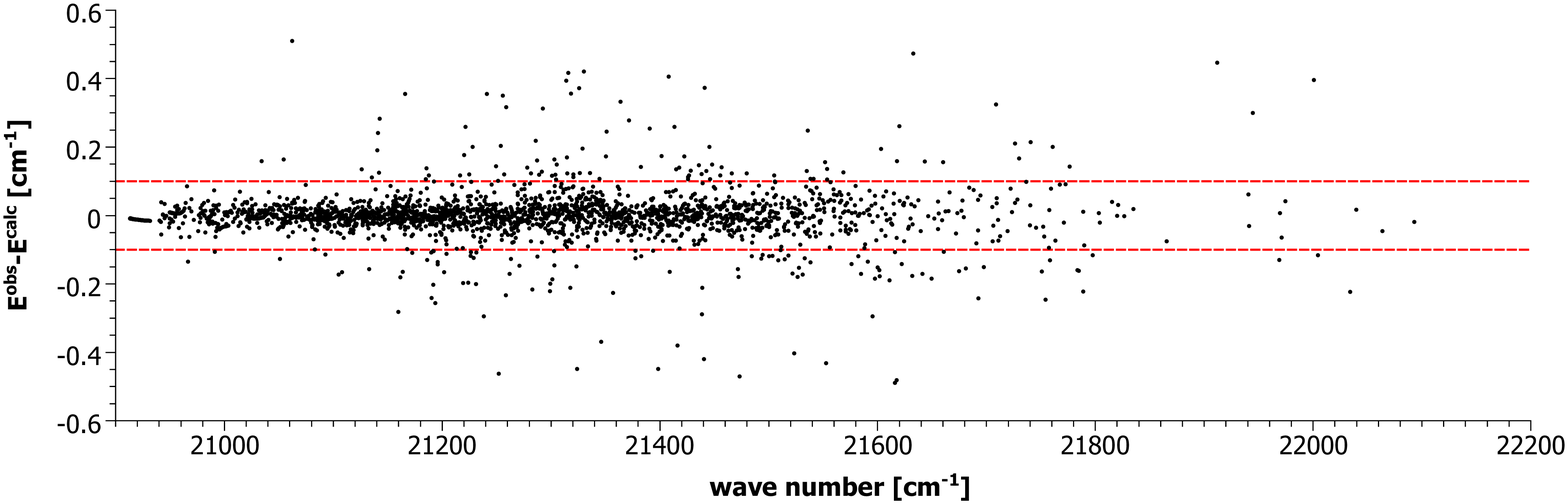}b
	\caption{Distribution of the residuals in line positions of the C(2)$^{1}\Pi_{u}$ $\leftarrow$ X$^{1}\Sigma^{+}_{g}$ system in Rb$_2$ before (a) and after (b) the coupled-channels fit. The horizontal lines (red online) mark borders of the uncertainty band corresponding to the experimental uncertainty, i.e. to $\pm 0.1$~\rcm. Note a difference in the vertical axis scale of both panels.}
	\label{residuals}
\end{figure}

The model reproduces positions of 2577 energy levels (2035 from the PLS experiment and 542 from Refs.~\cite{Lee,Lee2}) with a root-mean-square (rms) deviation of 0.069~\rcm\ and a dimensionless standard deviation of 0.77. When applied to the 2950 PLS experimental frequencies, the rms deviation amounts to 0.075~\rcm. To assess the quality of the fit we provide a plot of the residuals (Fig.~\ref{residuals}b). The red horizontal dashed lines indicate the experimental uncertainty. For comparison we performed also a single channel fit of all data from the present experiment (i.e. excluding data on the 2\TPst\ and 3\TSst\ states taken from Refs.~\cite{Lee} and \cite{Lee2}) with a single model function, namely the PEC of the 2\SPst\ state. The residuals of this fit are shown in Fig.~\ref{residuals}a (part of them can be seen also in Fig.~\ref{1267}a). As expected, a significant number of levels is well outside the experimental uncertainty in this simplified model.

In Ref.~\cite{Lee} (Table IV) the authors list the spin-orbit constant $A_v$ which determines the shift of the 2$^3\Pi_0$ levels with respect to those of the 2$^3\Pi_1$ state. It is interesting to compare these values with matrix elements $\langle ^3\Pi, v'|A_{^3\Pi^3\Pi}(R)|^3\Pi, v'\rangle$ calculated using the radial function from the present study (see Table~\ref{Av}). The agreement is better for higher vibrational levels, while for lower ones the deviations become larger and reach 0.17~\rcm. Around the potential minimum, the present $A_{^3\Pi^3\Pi}(R)$ function is determined mainly by the same data as in Ref.~\cite{Lee}, so one would expect better agreement. However, in Ref.~\cite{Lee} the whole shift was attributed to the diagonal spin-orbit interaction whereas in the present study the 2\TPst\ state levels are shifted also by other interactions (see eqs.~(\ref{SOs}) and (\ref{rots})). Table~\ref{Av} shows that the diagonal spin-orbit contribution is indeed dominant, but for $v'=0-5$ about 10\% of the shift has probably a different origin.

\begin{table}\fontsize{8pt}{13pt}\selectfont
\caption{Comparison between the spin-orbit constant $A_v$ (in \rcm) which determines shift of the 2$^3\Pi_0$ state levels with respect to those of the 2$^3\Pi_1$ state, as found in Ref.~\cite{Lee} and determined in this study.}
\begin{tabular}{|c|ccccccccccc|}\hline
$v$&0&1&2&3&4&5&6&7&8&9&10\\\hline
Ref. \cite{Lee} &1.54&1.61&1.71&1.82&1.94&2.03&2.14&2.27&2.37&2.52&2.66\\\hline
This study&1.37&1.51&1.64&1.77&1.88&2.00&2.13&2.26&2.40&2.54&2.69\\\hline
\end{tabular}
\label{Av}
\end{table}

\section{Conclusions}

In this paper we presented an example of deperturbation, where most part of the experimental data belongs predominantly to one of the interacting electronic states (C(2)\SPst) and the other states are visible mainly through the interactions with this `bright' state. This happens because the off-diagonal matrix elements of the molecular Hamiltonian are relatively small and significant mixing appears only in case of small separation between levels. The perturbations could be classified as local, but in fact they are so numerous that regular perturbation patterns like displayed in Figure~\ref{1267}a are rather rare. The problem seems underdetermined since finally to model the experimental observations four potential curves and seven radial functions of the perturbing operators had to be deduced. It was not straightforward to build the final model and to fit the experimental data with physically sensible functions, but this goal has been achieved. Apart from our data we used for this purpose the existing experimental data on the 2\TPst\ and 3\TSst\ states as well as the high quality theoretical predictions \cite{stany5-1, Pazyuk:2015}.

Although the residuals of the fit look convincing, we should note that there are still few dozens of lines, frequencies of which deviate from the model predictions by more than 3 standard deviations ($>0.3$ \rcm, see Fig.~\ref{residuals}b). We could not incorporate them into the list of `good' lines and so the question whether their assignment is wrong or the model needs refinement is left open. Nevertheless, it should be stressed that the presented model allowed to reproduce successfully positions of nearly three thousands lines in the very complex spectrum.

\section{Acknowledgements}

AP acknowledges partial support from the Sofia University Grant 80-10-44/2022 and Grant BG05M2OP001-1.002-0019 ``Clean technologies for sustainable environment -- waters, waste, energy for circular economy'', financed by the Operational programme ``Science and Education for Smart Growth 2014-2020'', co-financed by the European union through the European structural and investment funds. PK, JS and WJ were partially supported by the National Science Centre of Poland (Grant No. 2016/21/B/ST2/02190).

\section{Appendix A. Supplementary data}

Supplementary data associated with this article can be found  as at the address http://dimer.ifpan.edu.pl.

\end{document}